\documentstyle[aps,preprint,prl]{revtex}
\begin{document}

\title{\bf Fingerprints of Chaos }
\author{Virgil Baran \footnote{baran@lns.infn.it}
and Aldo Bonasera \footnote{bonasera@lns.infn.it}}

\address{\it  
Laboratorio Nazionale del Sud - Istituto Nazionale di Fisica Nucleare,
via S. Sofia 44, I-95123 Catania, Italy    }

\maketitle

\begin{abstract}
The asymptotic distance between 
trajectories $d_{\infty}$, is studied in detail to 
characterize the occurrence of chaos.  We show that this quantity is 
quite distinct and complementary to the Lyapunov exponents, and it allows
for a quantitave estimate for the folding mechanism which keeps the motion
bounded in phase space.  We study
the behaviour of $d_{\infty}$ in simple unidimensional maps.  Near a 
critical point $d_{\infty}$ has a power law dependence on the control
parameter.  Furthermore, at variance with the Lyapunov exponents, it
shows jumps when there are sudden changes on the available phase-space.
\end{abstract}
 
\newpage

\noindent

  One of the most characteristic feature which is emerging when dealing
with nonlinear systems is the appearance
of chaotic motion. There has been a considerable amount of work to 
establish what are the conditions for a nonlinear system (dissipative or
conservative) to display chaotic dynamics and what are some suitable 
quantities to characterize it.  The following gives a brief review of the
quantities discussed so far in the literature \cite{pp1}:

i) The  Lyapunov exponent (LE), the mean rate of separation between two adjacent
trajectories in phase space,  is one of the most used measures.
 A system is chaotic
when the trajectories diverge exponentially, i.e. the LE is larger than zero. 

ii) The correlation function characterizes the "memory" along 
one trajectory.  It decays quickly to zero in the chaotic regime.

iii) The power spectrum changes from discrete lines to a broad-band noise 
when chaos sets in.

These quantities can be calculated from the nonlinear equations
 of motion which describe the system.  Often, however the equations of
motion are not known but there might be some experimental determination of the
time evolution of a physical quantity.  
There are in the literature some suggestions on
how to extract the LE from a time series \cite{wolf}.  However the proposed
methods are not unique and depend on some working parameters \cite{casa}.
Other cases of interest exist where there is
no information at all about the time evolution of some quantity and final
phase space distributions are only known.  For instance in the field of 
nucleus-nucleus
collisions there is an active search for a liquid to gas phase transition
at excitation energies of tens MeV/particle or a phase transition to a
 quark gluon plasma at much higher excitation energies.  In these experiments
the final momentum distributions (of almost all) particles are known.  There
is absolutely no information about the time evolution of the system and there
is no way to estimate the LE from data.  Clearly,
it becomes very important to have hold of some clearly defined physical 
quantity that could define unambiguously the occurrence of a phase transition.
It is the purpose of this paper to show that there is a quantity, the 
asymptotic distance between trajectories, which could be easily estimated in
such cases where the quantities (i-iii) cannot be unambiguously determined and
which gives for instance as much information as the LE alone.
In particular we will describe how the distance
between trajectories, which defines the LE, saturates
 because of the finiteness of the available phase space.  
  We would like to stress that
in the literature \cite{pp1,wolf} particular care has been taken to estimate
 the LE by avoiding such saturation of the trajectories.  Here, we are going to
do the opposite, i.e. study in detail the saturation properties of trajectories
to demonstrate how many physical informations can be obtained from such
studies.
We will discuss some illustrative examples using one-dimensional maps. 
 Preliminary studies for 
Hamiltonian systems can be found in \cite{our}.

A one-dimensional map can be written as:

\begin{equation}
x_{n+1} = f(x_{n},r)
\end{equation}
where $r$ is a control parameter. Such maps, even though apparently very simple,
exhibit most of the chaotic features found in more complex systems.
   
The appropriate quantity to discuss here is the mean distance between 
two trajectories separated initially by a very small distance $d_0$:

\begin{equation}
d_n = {1 \over N} \sum_{i=0}^{N} \mid x_{n}^{(i)} - x{'}_{n}^{(i)}\mid
\end{equation}

and:

\begin{eqnarray}
x_{n}^{(i)} &=& f^{n} (x_{0}^{(i)}) \\
x{'}_{n}^{(i)} &=& f^{n} (x_{0}^{(i)}+ d_{0} )
\end{eqnarray}

In order to avoid fluctuations arising from a particular choice of the
starting point for the iterations we average over N couples of trajectories
and choose the initial starting points $x_0^i$ randomly from a uniform 
distribution. After n iterations, 
for small enough initial separation, the approximate n dependence of the
distance $d_n$ is: 

\begin{equation}
d_{n} = e^{ n \lambda} d_0 = e^{\lambda} d_{n-1}  \label{dist1}
\equiv \Lambda d_{n-1}
\end{equation}
where $\Lambda=e^{\lambda}$ and $\lambda$ is the Lyapunov exponent of the map:

\begin{equation}         
\lambda =\lim_{n \rightarrow \infty} {1 \over n} {\sum_{i=1}^{i=n}
{ln \mid f'(x_n) \mid}}         
\end{equation}
the prime indicates the derivative of the function $f(x_n)$ respect to $x_n$
 \cite{pp1}.
The action of the maps consists generally of two steps: the
stretching, which leads to the exponential regime and is characterized
quantitatively by the LE and the folding process which keeps the orbit bounded.
Therefore when $n$ is
sufficiently large, the solution eq.(5) is no longer valid.
We can consider eq.(5) as a first order expansion in $d_n$, and in the 
hypothesis that: {\it$d_n < 1$ for any n} , 
we include a second order correction term
in (\ref{dist1}):

\begin{equation}
d_{n+1} = \Lambda d_n - \Gamma d_{n}^{2}=F(d_n)         \label{dist2}
\end{equation}

Note the analogies to the derivation of the logistic map \cite{deri}.

Let us define the asymptotic value of the mean distance between two 
trajectories as:

\begin{equation}
d_{\infty}=\lim_{n \rightarrow \infty}
{1 \over n} \sum_{i=1}^{n} d_i          
\end{equation}
The fixed points of (\ref{dist2}) are $d_1=0$  and:

\begin{equation}
d_2=d_{\infty} = { \Lambda - 1.\over \Gamma }                \label{diinf}    
\end{equation} 
Thus, (\ref{dist2}) describes the irreversible approach of the
system to equilibrium which correspond to the fixed points solution (9).
From the stability condition $\mid F'(d_i) \mid < 1$ \cite{pp1},
we find that $d_1$ is
a stable point for $\lambda <0$, while $d_2=d_\infty$ is a stable point
for $0< \lambda <ln(3)$.  We also stress
these particular and interesting cases:

a) $F'(d_{\infty})=0$ gives $\lambda=ln(2)$ which is a superstable point of
the map $F(d_n)$.  Notice that such value of the LE is obtained in the
triangular and logistic maps for a control parameter where the maps
become ergodic.  {\it Thus ergodicity of the maps is equivalent to a 
superstable point of our proposed application describing the
 evolution of the distance between 
trajectories}. Also, $d_1=0$ becomes a superstable point
for $\lambda \rightarrow -\infty$. 

b)$F'(d_{\infty})=-1$ gives $\lambda=ln(3)$ which is the value for which
the map $F(d_n)$ has a pitchfork bifurcation.  In this case the values of
 $d_{\infty}$ at bifurcation are given by the conditions $d_1=F(d_2)$ and
 $d_2=F(d_1)$.  This and larger value cases of 
the LE are outside the purpose of this paper and will be 
discussed in a future publication \cite{mpb97}.

The actual value of $\Gamma$ can be easily obtained inverting
(\ref{diinf}):

\begin{equation}
\Gamma =  {e^{\lambda} - 1. \over d_{\infty}}        \label{Gamma}
\end{equation}

 The entire 
evolution of the distance between trajectories is given
by equation (\ref{dist2}).  It contains three characteristic quantities,
${\lambda}$, $d_{\infty}$ and ${\Gamma}$, but only two of them are independent
because of the relation (\ref{Gamma}).  To better grasp the meaning of the
physical quantities introduced above let us consider the (unbound) map:
$x_{n+1}=2x_n$.  It easy to show that for this map it is ${\lambda}=ln 2$,
and ${\Gamma}=0$.  If we impose the condition on the map to have modulus 1
(Bernoulli shift), the LE remains the same while ${\Gamma}=1/d_{\infty}=3$.
Thus the LE is only sensitive to the stretching mechanism while ${\Gamma}$ is
sensitive to the folding and stretching mechanisms, eq.(\ref{Gamma}): {\it its
knowledge allows us to distinguish between maps which have the same LE}.  

In fig.(1) we  plot $d_n$ versus n 
as obtained numerically for the logistic, triangular and sin maps, for three
different initial values of $d_0$\cite{maps}, full lines.
We observe in all cases
 that, after a fast increase, the distance between trajectories saturates.
The value of saturation is $d_{\infty}$ as defined in (8) independent 
on the initial relative distance $d_0$.  Inserting the values
of ${\lambda}$ and $d_{\infty}$, as obtained from eqs.(6) and (8), in eqs. (7)
and (10), gives the dot points displayed in fig.(1).  The agreement
to the numerical results is extremely good for all cases and
supporting our hypothesis.
{\it We conclude that in order to have an overall description for the evolution
of the mean distance between trajectories, 
beyond the exponential regime cf. eq.(5),
 we need two parameters, $\lambda$ and
$d_{\infty}$, cf.eq.(7)}.

At the ergodic point, corresponding to fully developed chaos, we can calculate 
analytically the LE and 
$d_{\infty}$ as mean values over the phase space using the
corresponding distribution function $\cite{pp1}$.  We obtain:

\begin{eqnarray}
\lambda_e  =  \int_{0}^{1} \rho(x) ln \mid 4-8x \mid = ln2   \\ 
d_{\infty e} = \int_{0}^{1} \rho(x) \rho(y) \mid x - y \mid = {{4 \over \pi^2}} 
\end{eqnarray}
for the logistic map.  For the triangular map we have $\lambda_e= ln2$,
and $d_{\infty e}=1/3$, the subscript e stands for ergodic.
 These values are 
in perfect agreement with the ones used in fig.(1a), and fig.(1c).

For other values of the control parameter it is not possible to 
calculate 
analytically $\lambda$ and $d_{\infty}$ and  we have performed some numerical
calculations displayed in fig.(2).  Figs.(2a) and (2c) give
the LE and $d_{\infty}$ respectively 
as obtained from eqs.(6) and (8) for the logistic
map.  The right column refers to the results of the
triangular map.  Recall that the LE for the triangular map is given by
$\lambda=ln(2r)$.  
These are some features of particular interest:

i) at particular values of the control parameter, $d_{\infty}$ has jumps which 
indicate a change in the dynamical behavior. 
Notice, for instance, the jump
near r=0.7 for the triangular map.  This jump,not observed in
the LE, is due to
the sudden increase of the available phase space,
 because of band splitting bifurcation,
 see also fig.(2b);

ii) similarly to the LE, at the
transition to regular windows $d_{\infty}$ drops to zero.

Let us study this last point in more detail for the logistic map. 
Near the transition point 
from order to chaos, the LE behaves like
\begin{equation}
\lambda \propto (r - r_{\infty})^{\beta}~~~~~~~;~~~~~~~\beta =
{ln2 \over ln\delta}
\end{equation}
Here $r_{\infty}$ is the accumulation value of control parameter for the 
double period bifurcation cascade.
This relation can be easily obtained by means of the Renormalization Group
Theory (RGT)\cite{pp1}.  Within the same framework we can easily 
derive $d_{\infty}$ and obtain\cite{rgt}:

\begin{equation}
d_{\infty} \propto  (r - r_{\infty})^{\nu/2}~~~~~~~;~~~~~~~\nu/2 =
{ln\alpha \over ln\delta}
\end{equation}
In the inset of fig.(2c), we plot the numerical result for $d_{\infty}$ vs. r near
the critical point.  The full line gives the power law, eq.(14), in
very good agreement to the numerical values.  
We remark that the critical exponent for  $d_{\infty}$ depends on both
Feigenbaum constants $\alpha$ and $\delta$, while the LE depends on
$\delta$ only. We have used the symbol $\nu$
in analogy to the treatment of second order phase transition.  Infact, the 
counterpart of the distance between trajectories for maps is given by the
variance in momentum space for Hamiltonian systems \cite{our}.   Near a second order
phase transition such a variance is proportional to the inverse of the
correlation length which depends on $(T-T_c)^{-\nu}$, where $T_c$ is
the critical temperature \cite{pp8}.

To make our analogy to phase transitions more complete recall that at the
 critical point the correlation function, defined as\cite{pp1}:
\begin{equation}
C(m,f) = \lim_{n \rightarrow \infty} {1 \over n}
\sum_{i=1}^{n} y_i*y_{i+m}
\end{equation}
where $y_i=f^{i} (x_0)- x_{av}$ and 
$x_{av}= \lim_{n \rightarrow \infty} {1 \over n}
\sum_{i=1}^{n} x_i$,
decays with a power law in $m$:
\begin{equation}
C(m,f_{r_{c}}) \propto m^{- \eta}~~~~~~~;~~~~~~~ \eta = {2 ln \alpha \over ln2}
\end{equation}
which depend now only on $\alpha$.

All above three critical exponents can be related as follows:

\begin{equation}
{\nu} =  \beta \eta
\end{equation}
which could be considered valid for the transition to chaos
through double period bifurcation.
This behaviour is similar for all the
maps having period doubling chaotic bifurcation, the critical exponents
depending on the values of corresponding $\alpha$ and $\delta$ constants.

For completeness in fig.(3) we plot the parameter $\Gamma$ vs. r for three
different maps \cite{maps}. We see that this parameter is very large 
especially near the critical point for the transition from
order to chaos, and then it becomes almost constant.  The value of the
constant is ${1 \over d_{\infty e}}$, eq.(12). 
 Thus the LE is to a good approximation proportional to $d_{\infty}$
 especially near the ergodic point.  This fact is especially 
important for equilibrated physical systems since one can deduce the
 properties of the LE from the variances which are given from thermodynamics
\cite{our}.

In conclusion, we have shown that a 
useful quantity to characterize the occurrence of chaos is
$d_{\infty}$, the value at which the distance between two trajectories
saturates  in phase space.  
It is complementary to the Lyapunov exponent giving informations 
about the global features of phase space. It signals also 
when changes in dynamics are having place
which are not always reflected in the behavior of the Lyapunov exponent.     
Both parameters {\it are needed} in order to obtain a correct description
of the time evolution for the mean distance between trajectories. 
In analogy to phase transitions, the LE is the order parameter while 
$d_{\infty}$ is the inverse of the correlation length.
In some experiments both quantities can be measured, often, however, like
in nucleus nucleus or cluster cluster collisions, it
is not possible to follow the time evolution of the system and asymptotic
quantities, like $d_{\infty}$, can only be detected.
The generalized application for the distance between trajectories proposed
in this paper, eq.(7),
gives a link between the initial exponential expansion and the final
equilibrium stage.  We have shown that ergodicity is simply a 
superstable point of our proposed eq.(7) and naturally gives a Lyapunov
exponent  $\lambda= ln2$.

\bigskip
\bigskip
\centerline{\bf Acknolewdgments}
 One of the authors (V.B.) thanks the Istituto Nazionale di Fisica 
Nucleare-Italy for a postdoctoral fellowship.  A.B. thanks M.
Ploszajczak for stimulating discussions which inspired this work.

\newpage

\section*{ Figure captions }

\begin{description}

\item[Fig. 1] Distance between trajectories vs. iteration for different maps.
(a) Logistic map with the control parameter r=4; (b) the same as (a)
 but for r=3.771;
(c) triangular map at r=1;(d) sine map at r=0.73.
 The different curves correspond
to different starting values of $d_0$. The dots are obtained from eq.(7).

\item[Fig. 2] Lyapunov exponents and asymptotic distances between trajectories
vs. control parameter r,for the logistic map (left column) and triangular map
(right column).
Inset(2c): $d_{\infty}$ vs. r near the critical point for order to chaos
transition, for the logistic map .   The
full lines give the power law dependence of $d_{\infty}$ as discussed in the 
text.  

\item[Fig.3] $\Gamma$ vs. r for the logistic (a), sine(b) and triangular
map(c).  The dashed line gives the values of $\Gamma$ when $\lambda =ln(2)$ 
for the three maps respectively.
 
\end{description}


\begin{thebibliography}{99}
\bibitem{maps} The maps used in this paper are [1]:

a) logistic map:  $f(x_n)=r x_n(x_n-1)$, $0 \le r \le 4$;

b) triangular map: $f(x_n)=r(1-2 \mid {1 \over 2}-x_n \mid)$, $0 \le r \le 1$;

c) sine map:  $f(x_n)=x_n+r sin(2. \pi x_n)$, $0 \le r \le 0.7326$.
\bibitem{rgt} Introducing the doubling operator T [1]
and using eqs.(2-4) and (8),
we easily get: $d_{\infty}(Tf)=\alpha d_{\infty}(f)$.  Which can be iterated
to:  $d_{\infty}(f)=\alpha^{-n} d_{\infty}(T^{-n}f)$.  Following the same steps
as for the derivation of the power law dependence of the LE we obtain eq.(14).
\end{thebibliography}
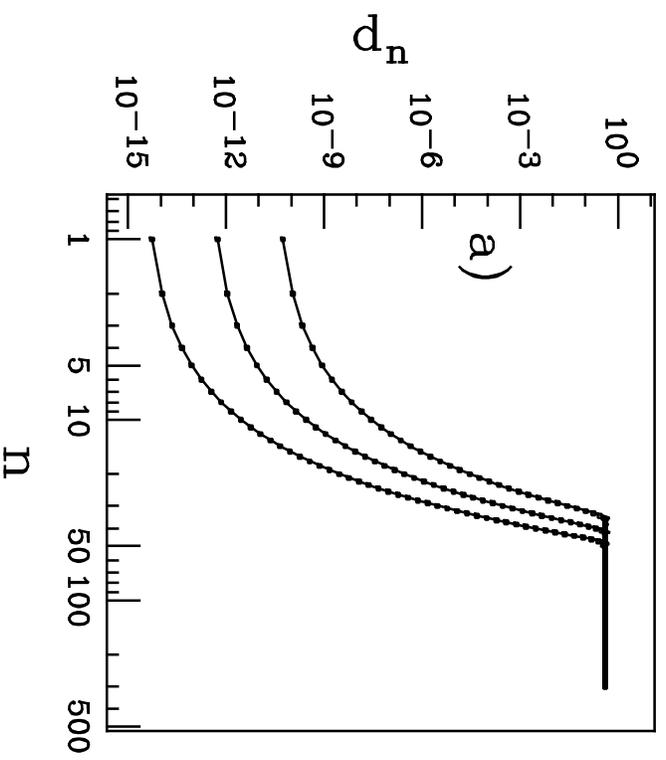
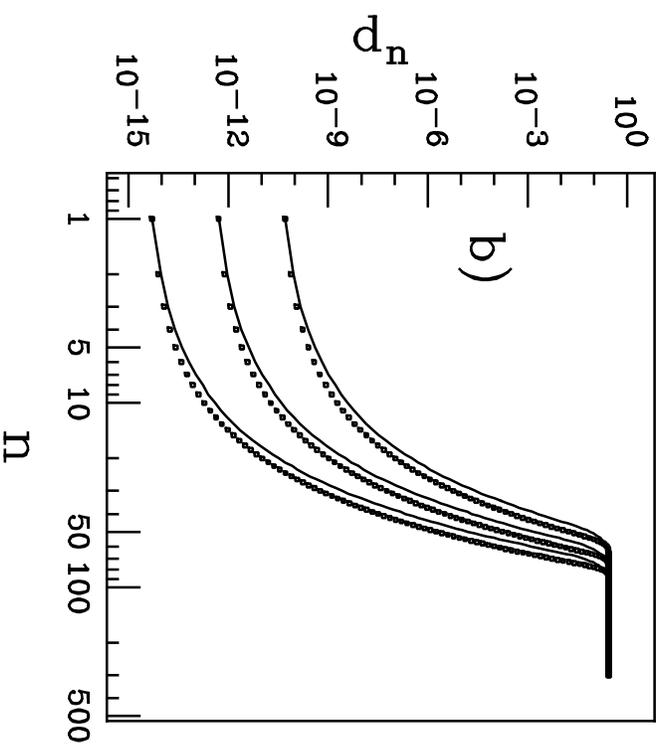
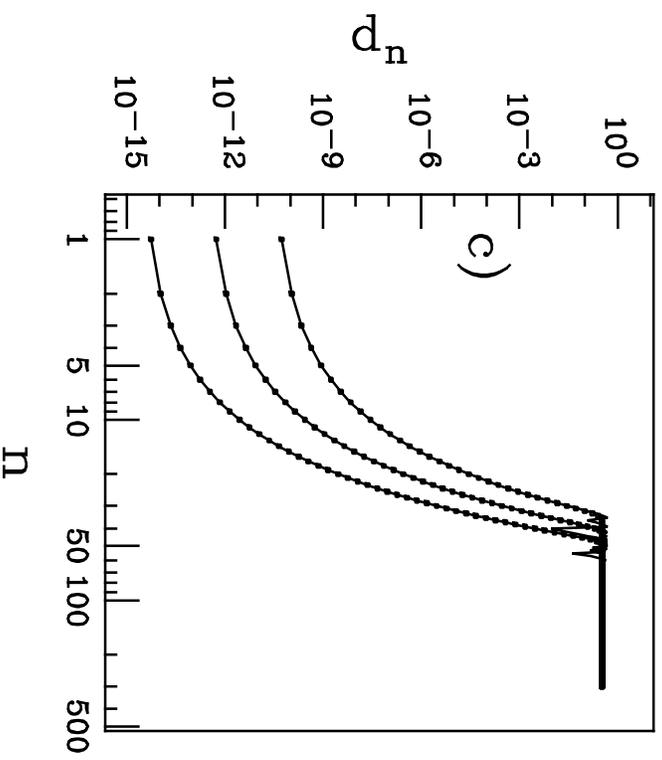
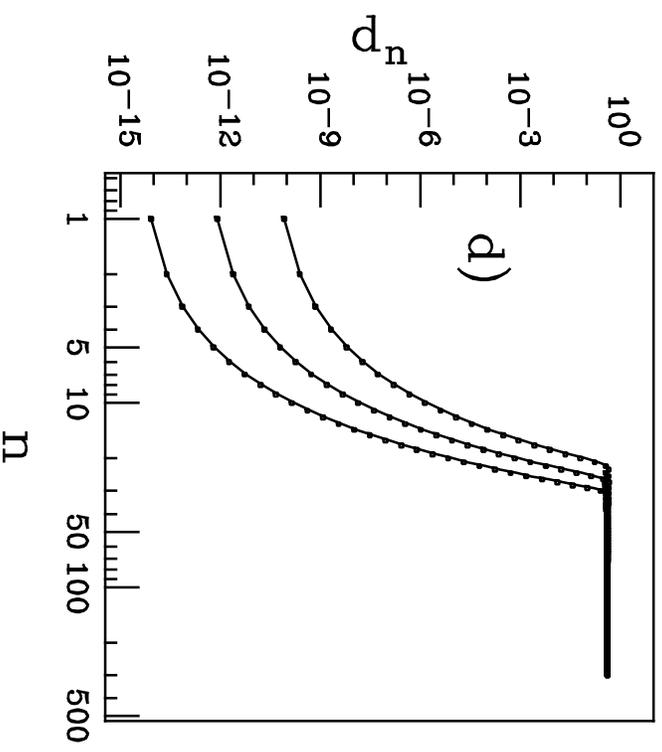
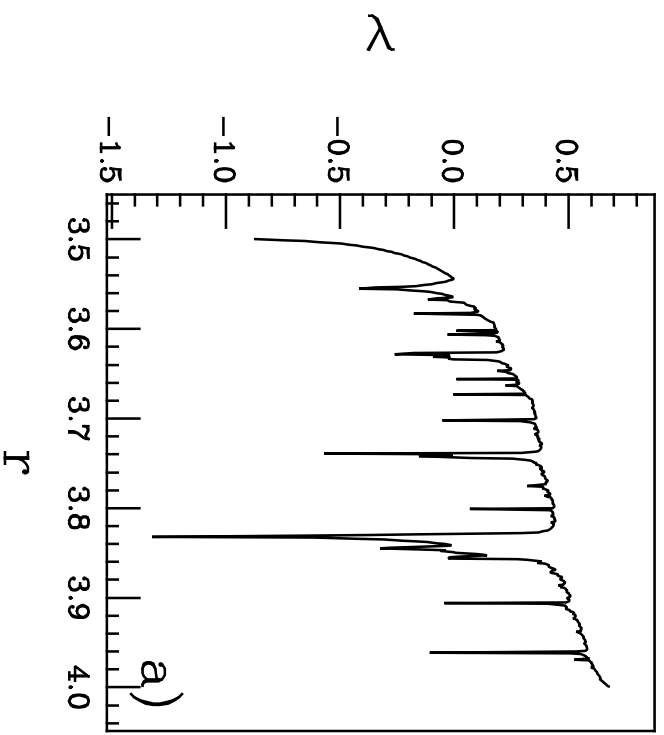
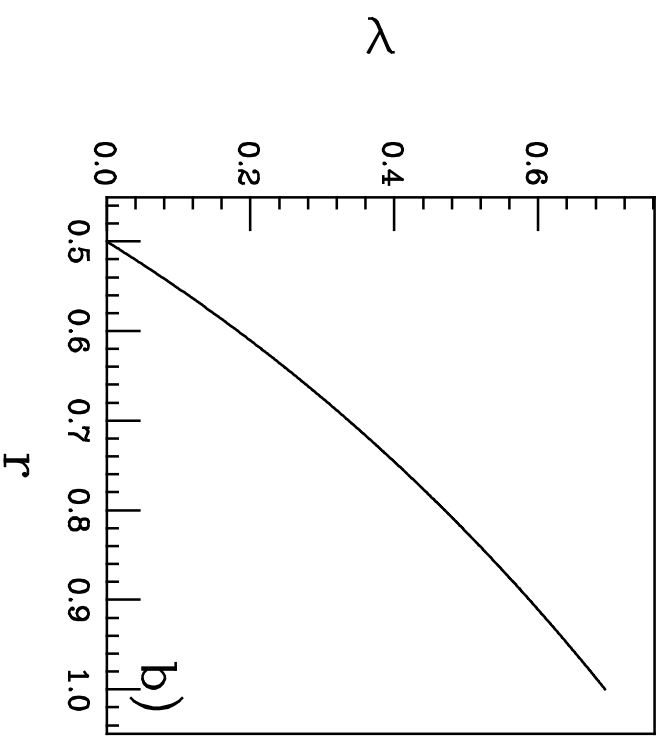
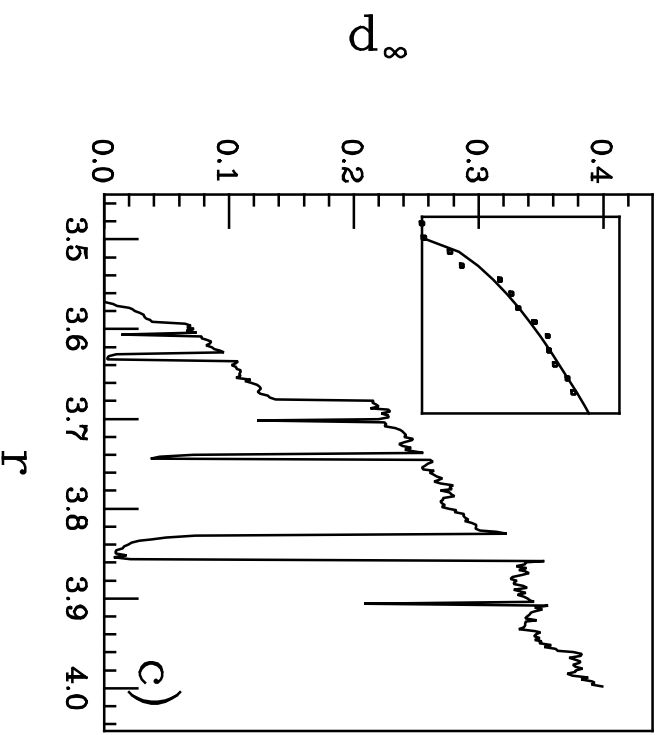
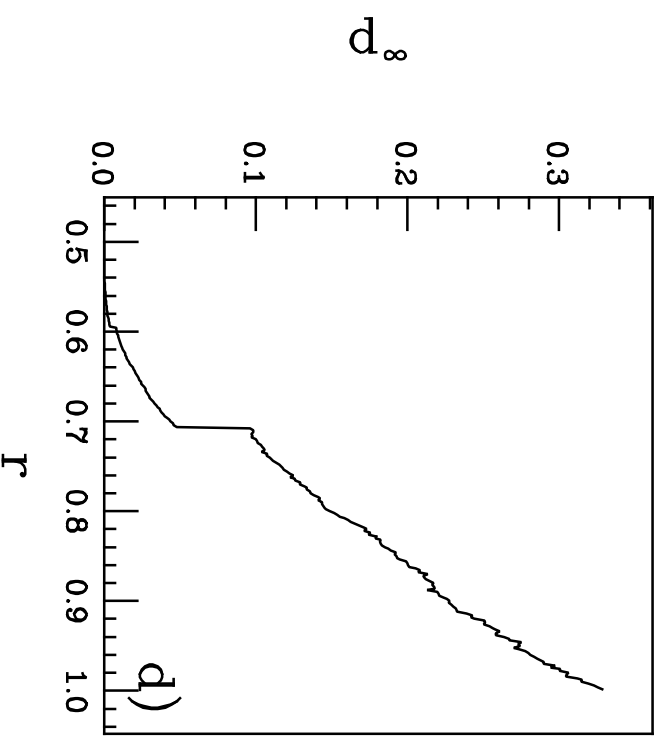
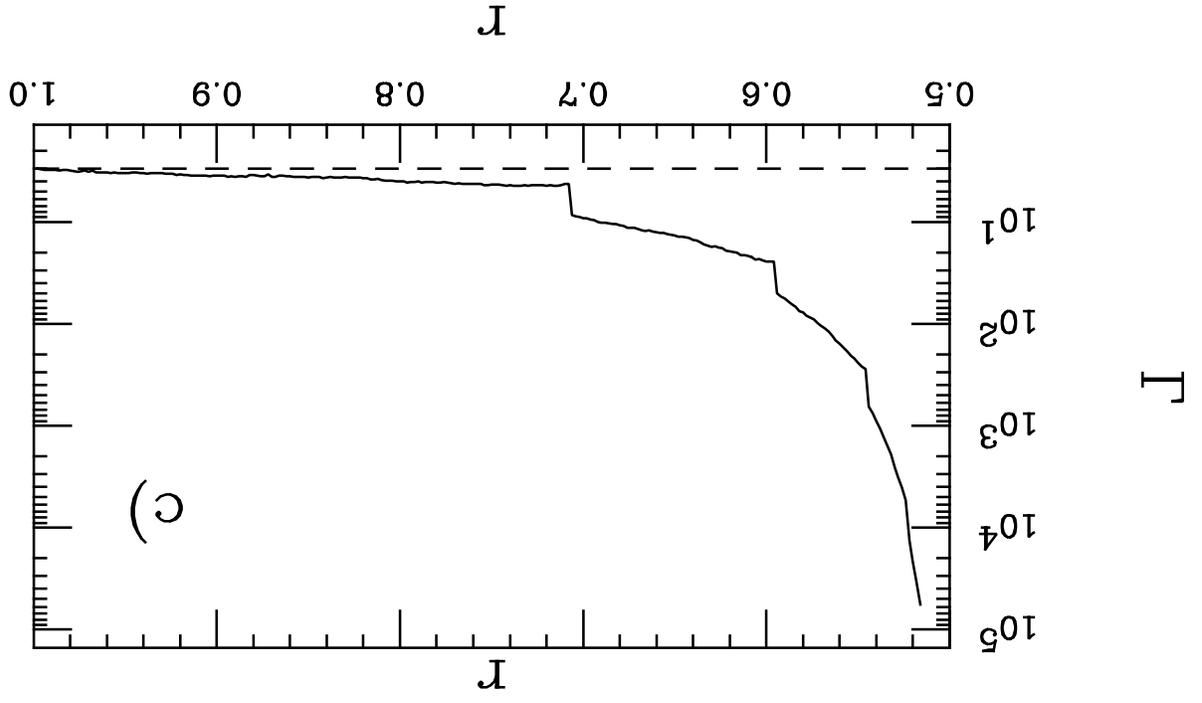
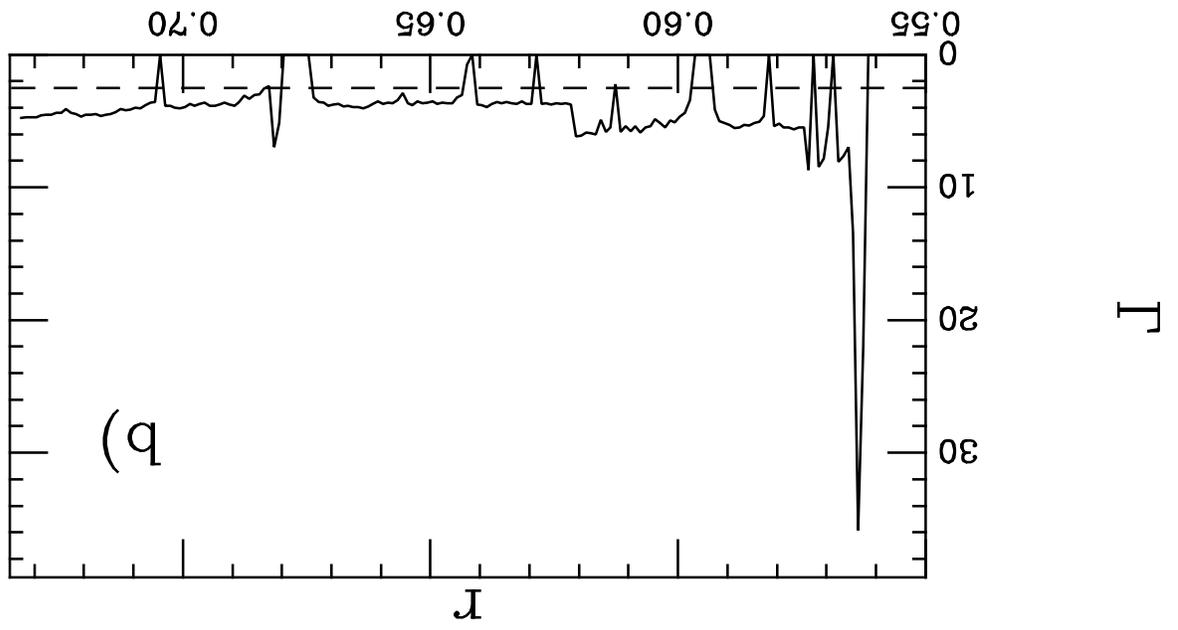
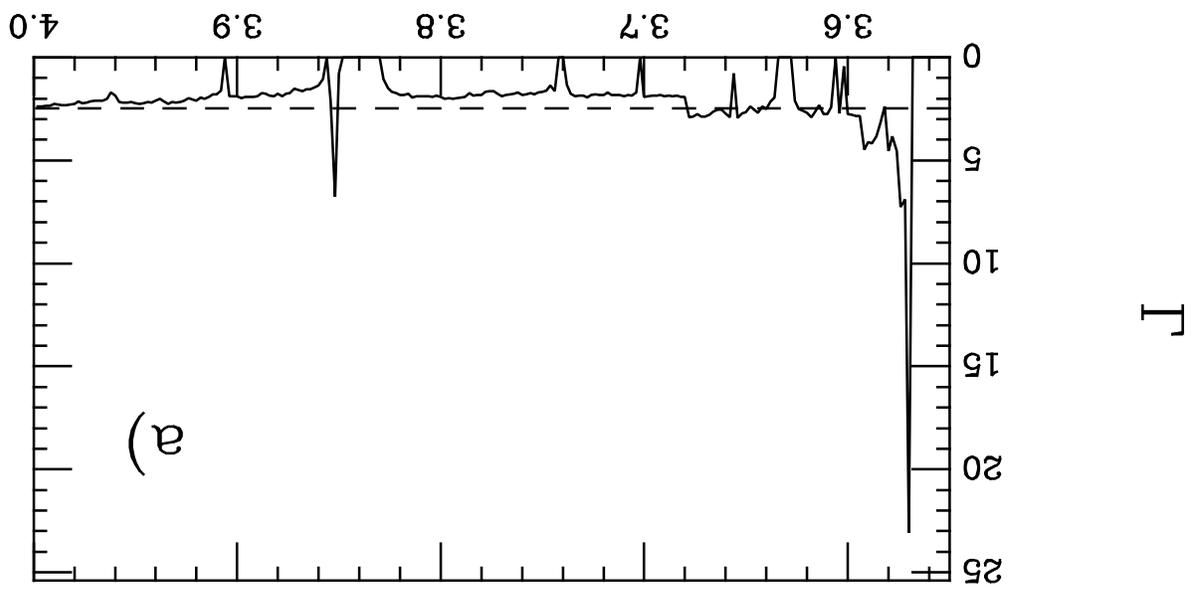
\end{document}